\documentclass[pdflatex,sn-mathphys-num]{sn-jnl}
\usepackage{amsmath,amssymb,amsfonts}%
\usepackage{amsthm}%
\usepackage{algorithm}%
\usepackage{algorithmicx}%
\usepackage{algpseudocode}%
\usepackage{listings}%
\usepackage{multirow}%
\usepackage{booktabs}%
\usepackage{rotating}
\usepackage[utf8]{inputenc}
\usepackage{enumitem}
\usepackage{subfig}
\usepackage{url}

\usepackage{graphicx}
\usepackage{caption}
\usepackage{listings}
\usepackage{xcolor}
\colorlet{punct}{red!60!black}
\definecolor{background}{HTML}{EEEEEE}
\definecolor{delim}{RGB}{20,105,176}
\colorlet{numb}{magenta!60!black}

\lstdefinelanguage{json}{
    basicstyle=\scriptsize\ttfamily,
    numbers=left,
    numberstyle=\scriptsize,
    stepnumber=1,
    numbersep=8pt,
    showstringspaces=false,
    breaklines=true,
    frame=lines,
    backgroundcolor=\color{background},
    literate=
     *{0}{{{\color{numb}0}}}{1}
      {1}{{{\color{numb}1}}}{1}
      {2}{{{\color{numb}2}}}{1}
      {3}{{{\color{numb}3}}}{1}
      {4}{{{\color{numb}4}}}{1}
      {5}{{{\color{numb}5}}}{1}
      {6}{{{\color{numb}6}}}{1}
      {7}{{{\color{numb}7}}}{1}
      {8}{{{\color{numb}8}}}{1}
      {9}{{{\color{numb}9}}}{1}
      {:}{{{\color{punct}{:}}}}{1}
      {,}{{{\color{punct}{,}}}}{1}
      {\{}{{{\color{delim}{\{}}}}{1}
      {\}}{{{\color{delim}{\}}}}}{1}
      {[}{{{\color{delim}{[}}}}{1}
      {]}{{{\color{delim}{]}}}}{1},
}

\usepackage{researchpack}
\algrenewcommand{\algorithmicrequire}{\textbf{Precondition:}}
\algrenewcommand{\algorithmicensure}{\textbf{Output:}}

\newtheorem{postulate}{Postulate}

\newcommand{\system}[1]{#1}
\newcommand{\systempop}[2]{#1-$\text{P}_{#2}$}
\newcommand{\systemfair}[2]{#1-$\text{F}_{#2}$}
\newcommand{\mosaic}[3]{#1-$\text{P}_{#2}\,\text{F}_{#3}$}

\begin{document}

\title[MOSAIC]{\textbf{MOSAIC: Multimodal Multistakeholder-aware Visual Art Recommendation}}

\author*[1]{\fnm{Bereket A. } \sur{Yilma}}\email{bereket.yilma@uni.lu}

\author[1]{\fnm{Luis A.} \sur{ Leiva}}\email{luis.leiva@uni.lu}

\affil[1]{\orgdiv{Department of Computer Science}, \orgname{University of Luxembourg}, \orgaddress{\street{6, avenue de la Fonte}, \city{Belval}, \postcode{4364}, \state{Esch-sur-Alzette}, \country{Luxembourg}}}

%%==================================%%
%% Abstract %%
%%==================================%%

\abstract{Visual art (VA) recommendation is complex, as it has to consider the interests of users (e.g. museum visitors) and other stakeholders (e.g. museum curators). We study how to effectively account for key stakeholders in VA recommendations while also considering user-centred measures such as novelty, serendipity, and diversity. We propose MOSAIC, a novel multimodal multistakeholder-aware approach using state-of-the-art CLIP and BLIP backbone architectures and two joint optimisation objectives: popularity and representative selection of paintings across different categories. We conducted an offline evaluation using preferences elicited from 213 users followed by a user study with 100 crowdworkers. We found a strong effect of popularity, which was positively perceived by users, and a minimal effect of representativeness. MOSAIC's impact extends beyond visitors, benefiting various art stakeholders. Its user-centric approach has broader applicability, offering advancements for content recommendation across domains that require considering multiple stakeholders.}


\maketitle

%% =============================================================================
\section{Introduction}
\label{sec:intro}
%% =============================================================================

Appreciating visual art (VA) is a complex and subjective experience that is influenced by a range of factors, including personal preferences, cultural context, and historical significance. Paintings are important VA items that bring together complex elements such as drawings, narration, composition, or abstraction~\cite{1130282273074769280}. As VA becomes more diverse and accessible, there is an increasing need for innovative approaches to facilitate art discovery, promote engagement, and ultimately support user satisfaction. This is challenging, since the emotional and cognitive responses triggered by VA can vary widely among users~\cite{naudet2015museum}. In this context, VA recommender systems (VA RecSys) have emerged as a promising solution to these challenges, as they can provide personalised recommendations based on users' preferences and interests~\cite{10.1145/3450613.3456847}. 

Overall, current approaches in VA RecSys essentially try to leverage different representations of paintings to identify similarities with user preferences and derive personalised recommendations~\cite{li2023siamese, wang2023research}. Unlike other domains considering e.g. our cultural heritage, the development of VA RecSys poses additional unique challenges. This originates from the growing need to factor in the diverse perspectives and expectations of stakeholders involved in the art ecosystem (such as artists, curators, collectors, museums, art enthusiasts, and educators) to generate suitable recommendations~\cite{nobre2022cultural}. The underlining assumption is that optimising only for previously elicited user preferences overlooks the diverse and evolving nature of human taste~\cite{palumbo2023visual}. However, the very essence of art is rooted in the belief that humans are learning creatures on a journey of exploration and discovery, picking up preferences and developing interests as they come across different content~\cite{zhitomirsky2023they}. 

To truly capture the essence of human experience towards art, VA RecSys must be reimagined, such that it reflects the natural, evolving nature of human preferences. As such, it is crucial for VA RecSys to expose users to new content and perspectives, rather than merely relying on a narrow set of assumptions regarding user preferences, which could be a disservice to the complex and diverse nature of art~\cite{ 10.1145/3450613.3456847}. Nonetheless, such a task for a VA RecSys is far from trivial. It calls for algorithms and techniques that can consider novel and diverse perspectives while striking a reasonable balance between already established user preferences, while also allowing users to control the level of exposition and exploration of new VA content. 

Early work on VA RecSys has primarily relied on inferring similarities and relationships among paintings based on high-level features derived from traditional metadata such as artist names, styles, and materials \cite{messina2017exploring, zhitomirsky2023they}. Nonetheless, these features are not sufficiently expressive to cater for the subjective taste of users. In recent years, researchers have started to focus on more advanced VA representation learning techniques to better capture semantic relatedness of art works. For example, He et al.~\cite{he2016vista} and Messina et al.~\cite{messina2019content} demonstrated that latent visual features extracted using Deep Neural Networks outperform textual metadata. Other studies have also argued in favor of visual features over textual metadata in VA RecSys~\cite{messina2017exploring, 10.1145/3450613.3456847}. 

Recent research has shown that both text and visual modalities capture the complex semantics embedded in VA, but it is their combination, using either late or early fusion techniques, the approach that delivers better performance~\cite{Yilmaleiva23}. A late fusion strategy combines rankings generated from independently trained modes, 
whereas early fusion, using multimodal representation learning techniques, consider a joint embedding from visual and textual data. Early fusion approaches have shown remarkable success in various down stream tasks~\cite{lialin2023scalable}, 
including in VA RecSys~\cite{YilmaleivaUMAP23}.

Contrastive Language-Image Pre-training (CLIP)~\cite{radford2021learning} 
and Bootstrapping Language-Image Pre-training (BLIP)~\cite{li2022blip} 
are the two most prominent early fusion approaches that have been used recently for VA RecSys~\cite{YilmaleivaUMAP23}.  
These multimodal models have also been successfully applied in several downstream tasks such as scene understanding~\cite{chen2023clip2scene}, image retrieval~\cite{sain2023clip}, or question answering~\cite{zhu2023chatgpt}. In VA RecSys, the learned representations can be leveraged to match elicited preferences from users and derive recommendations that are presumably accurate. However, in addition to finding efficient representations, seamlessly blending multiple stakeholder perspectives with user preferences remains an open challenge in VA RecSys. 
Hence, in this work we address the following research question: 
\textbf{How to effectively account for multiple stakeholders in user-centric VA recommendations?}
 
To answer this research question, we propose MOSAIC (MultimOdal multiStakeholder-Aware vIsual art reCommendation). 
We build multimodal representations of paintings by jointly training on textual descriptions and images of paintings and then propose different strategies to jointly optimise for multistakeholder objectives 
that promote user-centric beyond-accuracy measures~\cite{Kaminskas16} in VA recommendations.
In sum, this paper makes the following contributions:
\begin{itemize}
    \item We introduce MOSAIC, a novel VA RecSys approach that optimises for multiple stakeholders in the art ecosystem.
    \item We develop and study six variants of MOSAIC on top of two backbone architectures,
        to progressively and seamlessly introduce joint optimisation objectives.
    \item We conduct an offline evaluation to assess the impact of such joint objectives 
        and then conduct a user study to assess MOSAIC's performance from a user-centric perspective.
    \item We contextualise our findings and provide valuable insights on how to design next-generation VA RecSys blending learning, discovery, and user satisfaction.
        
\end{itemize}

%% =============================================================================
\section{Related Work}
\label{sec:rl}
%% =============================================================================

The concept of multistakeholder recommendation has gained significant attention recently, 
with the growing need for an inclusive approach that expands outward from the user, 
in order to include the perspectives and utilities of multiple stakeholders~\cite{abdollahpouri2020multistakeholder}. 
Historically, the RecSys community has been focused on creating value for consumers. 
However, these systems in practice serve some business or organisational objectives as well. 
Thus, recently more research has been devoted to properly consider the needs of all stakeholders 
involved by viewing recommendation as a problem that requires balancing the interests and objectives of multiple parties,
taking into account different optimization time horizons~\cite{surer2018multistakeholder}. 

In a multistakeholder environment, an ideal recommender system should aim to balance the objectives of all stakeholders to ensure a satisfactory outcome for everyone involved, including consumers, providers, and the recommendation platform, as well as potentially relevant side stakeholders such as society~\cite{abdollahpouri2021multistakeholder}. 
Many RecSys studies operate under the assumption that improving the accuracy of recommendations will ultimately lead to more valuable systems for users. It is a logical assumption that a better algorithm should present more relevant items at the top of the list, resulting in a better user experience. 
However, this assumption does not always hold, as just being accurate is not enough for a recommender system to be successful~\cite{mcnee2006being}. Therefore, while accuracy is undoubtedly an essential factor, it is not the only one.
A good recommender system must consider other user-related aspects that go beyond accuracy~\cite{shi2013trading}. 

Such observations have led to a strand of research on “beyond-accuracy” measures like diversity, novelty, or serendipity~\cite{isufi2021accuracy}. The domain of VA RecSys is among the fields where the notion of multistakeholder-awareness is vital yet often overlooked~\cite{10.1007/978-3-319-67162-8_35}. This is primarily due to the fact that paintings are both high dimensional and semantically complex, leading to a diverse and subjective set of emotional and cognitive reflections on users~\cite{wu2020paintkg, naudet2015museum}. Indeed, art paintings reflect the complex and intricate interplay of concepts ranging from individual ideas and beliefs to concepts with cultural and historical significance of a society~\cite{palumbo2023visual}. Furthermore, in the VA ecosystem there are multiple stakeholders that are directly or indirectly affected by the way recommendations are delivered to visitors such as artists, curators, collectors, museums, art enthusiasts, educators as well as society at large~\cite{nobre2022cultural}. 

The recent VA RecSys literature has been mostly focused on finding efficient data representation techniques that are capable of capturing the complex semantics embedded in paintings~\cite{messina2017exploring, messina2020curatornet, he2016vista}. A recently notable work by Yilma et al.~\cite{Yilmaleiva23} demonstrated that the key explanatory factors for semantic relatedness of art works can be captured from the fusion of textual and visual modalities. That work leveraged a late rank fusion strategy that combined rankings of recommendations generated from models independently trained on image and textual data. More recently, it has been shown that, instead of late rank fusion, jointly learning embedding from textual and visual data (early fusion) outperforms the late rank fusion strategy~\cite{YilmaleivaUMAP23}. This later work underlines the fact that the elements of visual art recommendation lie at the intersection of textual and visual modalities. Although efficient multimodal representations to capture abstract semantics in VA are important steps towards quality VA recommendations, it is still unclear how to factor in multistakeholder perspectives and generate VA recommendations that strike an optimal balance between accuracy, novelty, serendipity, and diversity. Thus, research effort to seamlessly incorporate learning and discovery with user preferences in VA RecSys is still considered a worthwhile endeavour, especially with regards to evaluating the quality of the recommendations from a user-centric perspective~\cite{han2017survey}.

%% =============================================================================
\section{Multimodal Visual art representation learning}
\label{sec:mmorl}
%% =============================================================================
Traditionally, late fusion (i.e. at post-hoc) has been preferred over early fusion (i.e. at the feature level) 
because the fused models can be independent of each other, so each can use their own features, numbers of dimensions, etc.~\cite{Ramachandram17}. However, recent advances in multimodal representation learning techniques have shown promising results in several downstream tasks~\cite{li2021align}, including VA RecSys~\cite{YilmaleivaUMAP23, yilma2024artful}.  
We now present two multimodal representation learning techniques based on the state-of-the-art CLIP and BLIP architectures,
and show how the learned representations can be utilised to generate VA recommendations. 

\subsection{Contrastive Language-Image Pre-training (CLIP)}
\label{subsubsec:clip}

CLIP is a technique for pre-training a language model on a large dataset of images and their associated text~\cite{radford2021learning}. 
Unlike traditional models that use an image encoder (e.g., CNN) and a classifier (e.g., a fully-connected network), 
CLIP jointly trains an image encoder and a text encoder to encourage a close embedding space 
between the ones that form a pair via contrastive learning. 
The model predicts the correct image given a text caption, and vice versa. 
It learns to generate a shared embedding space for both image and text inputs, 
where similar images and texts are close to each other in the embedding space. 
The idea is that by pre-training on this task, 
the model will learn to understand the relationship between language and visual concepts, 
which can then be fine-tuned for a variety of natural language understanding and image understanding tasks.  

Particularly, given a batch of $M <$image,text$>$ pairs, CLIP is trained to predict which of the $M$ $\times$ $M$ possible image-text pairings across a batch actually occurred. Thus, the model learns a multimodal embedding space by jointly training an image encoder, either a ResNet or a Vision Transformer (ViT), and a text encoder (a Transformer model, such as BERT) to maximise the cosine similarity of the image and text embeddings of the $M$ real pairs in the batch, while minimizing the cosine similarity of the embeddings of the $M^2$ - $M$ incorrect pairings. Hence, a symmetric cross-entropy loss is optimised over these similarity scores. 

\begin{figure}[!h]
\centering
\includegraphics[width= \textwidth]{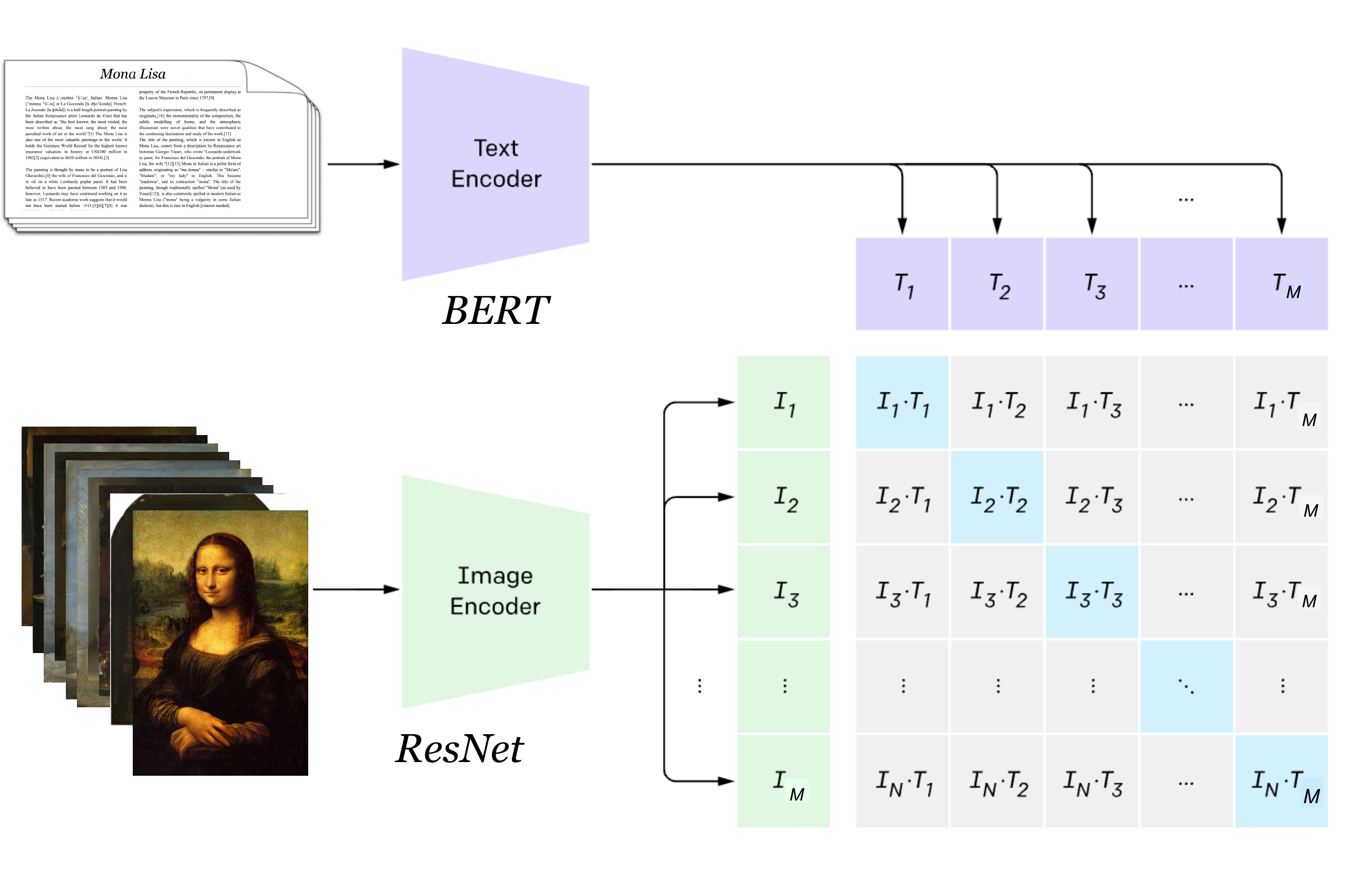}
\caption{Overview of our multimodal approach with CLIP to learn latent semantic representations of paintings.}
\label{fig:clip_feature_learning}
\end{figure}

In this work, we leveraged CLIP as a backbone feature extractor to learn a joint embedding space for images of paintings and their corresponding textual descriptions in order to uncover the latent semantic relationships between paintings embedded within the two modalities. We use ResNet-50 as our image encoder and BERT as our text encoder in our CLIP architecture. Once the features from both modalities are extracted, a matrix  $\mathbf{A} \in \mathbb{R}^{m \times m}$ is computed where each entry $\mathbf{A}_{ij}$ is the cosine similarity measure between the joint painting embeddings, which is then leveraged to compute semantic similarities of paintings for VA RecSys tasks. \autoref{fig:clip_feature_learning} illustrates our multimodal approach to learn latent semantic representations of paintings with CLIP. 

\subsection{Bootstrapping Language-Image Pre-training (BLIP)}
\label{subsubsec:blip}

BLIP is also a method for pre-training a neural network that combines language and image modalities~\cite{li2022blip}. 
However, unlike CLIP, BLIP pre-trains a model by learning to predict an image or a text given the other modality, 
with the goal of bootstrapping the model's understanding of multimodal relationships. 
BLIP introduces a mixture of encoder-decoder, a unified model which can operate in one of the following three functionalities. 
(1)~\emph{Unimodal encoder}, which separately encode image and text. The image encoder is a ViT and the text encoder is BERT. A \texttt{[CLS]} token is appended to the beginning of the text input to summarise the sentence. (2)~\emph{Image-grounded text encoder}, which injects visual information by inserting a cross-attention layer between the self-attention layer and the feed forward network for each transformer block of the text encoder. A special \texttt{[Encode]} token is appended to the text, and its output embedding is used as the multimodal representation of the image-text pair. (3)~\emph{Image-grounded text decoder}, which replaces the bi-directional self-attention layers in the text encoder with causal self-attention layers. A special \texttt{[Decode]} token is used to signal the beginning of a sequence.

BLIP jointly optimises three objectives, namely Image-Text Contrastive Loss (ITC), Image-Text Matching Loss (ITM), and Language Modeling Loss (LM). 
ITC activates the unimodal encoder with the aim of aligning the feature space of the visual and text transformers by encouraging positive image-text pairs to have similar representations in contrast to the negative pairs. 
ITM activates the image-grounded text encoder for a binary classification task, 
where the model is asked to predict whether an image-text pair is positive (matched) or negative (unmatched) given their multimodal feature. 
Finally, LM activates the image-grounded text decoder, which aims to generate textual descriptions conditioned on the images.

\begin{figure}[!h]
\centering
\includegraphics[width= \textwidth]{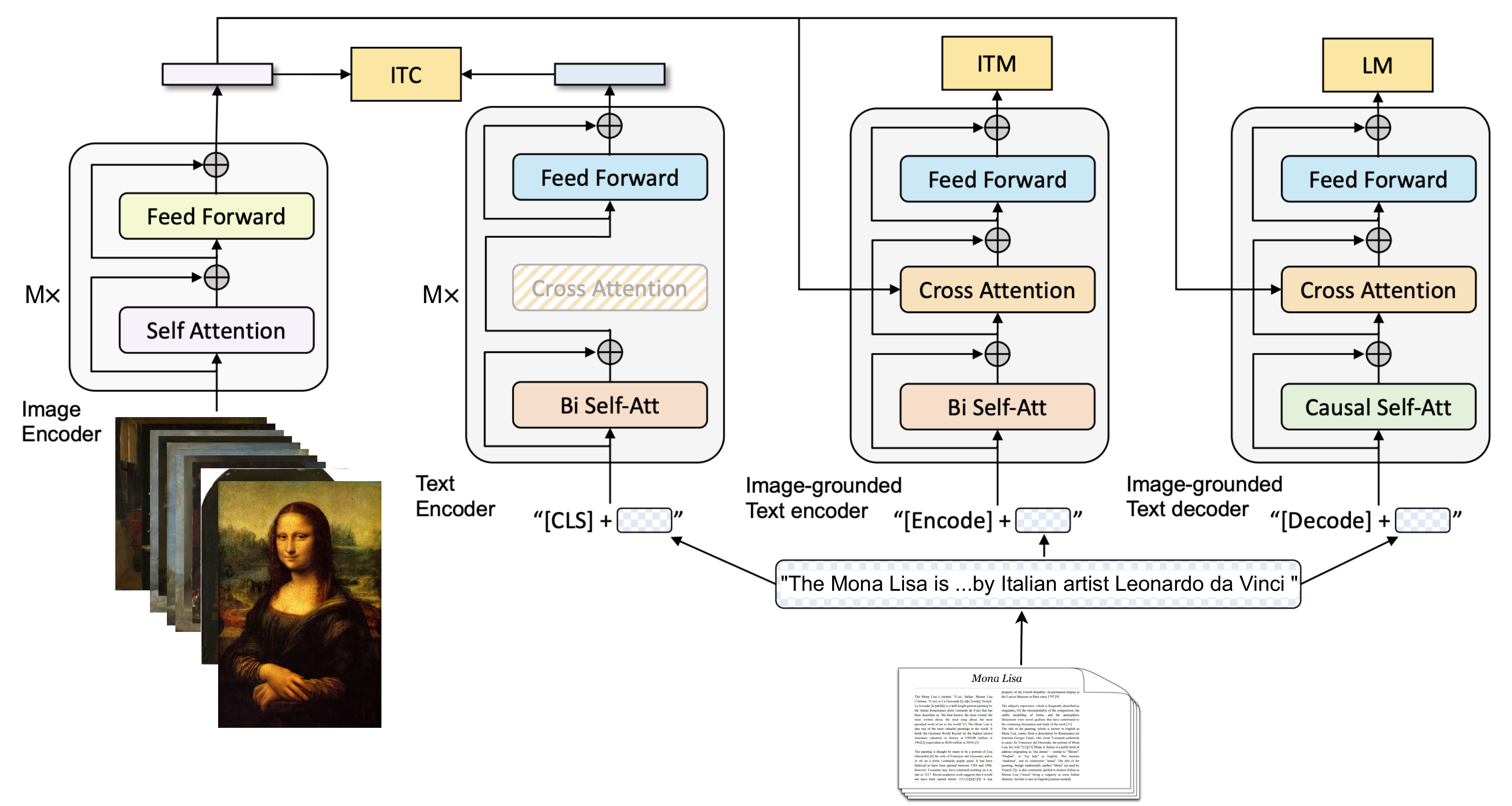}
\caption{Overview of our multimodal approach with BLIP to learn latent semantic representations of paintings.}
\label{fig:blip_feature_learning}
\end{figure}

For the task of VA representation learning, we leveraged the pre-trained BLIP model as a multimodal feature extractor. 
We compute ITM scores for every painting by first extracting multimodal features 
and then passing them to the ITM head, which generates probability matching (i.e., similarity) scores for each $<$image,text$>$ pairs. 
Afterwards, a matrix  $\mathbf{A} \in \mathbb{R}^{m \times m}$ is computed where each entry $\mathbf{A}_{ij}$ is the probability matching score between the joint painting embeddings which is then leveraged to compute semantic similarities of paintings for VA RecSys tasks. \autoref{fig:blip_feature_learning} illustrates our multimodal approach to learn latent semantic representations of paintings with BLIP.

%% =============================================================================
\section{Multistakeholder-aware Visual Art Recommendation}
\label{sec:method}
%% =============================================================================

In addition to the users of VA RecSys, 
the art ecosystem brings together different stakeholders that have a direct or indirect influence on users. 
The two most important stakeholders in VA recSys are the crowd, who contribute to deciding the popularity of art pieces, and museum curators, who aim to promote certain topics, themes, artists, and stories. 
Concurrently optimising for these two objectives also reflects a subset of stakeholders, including artists, collectors, and society at large~\cite{10.1007/978-3-319-67162-8_35, nobre2022cultural}. Hence, VA RecSys needs to consider such relevant stakeholders to blend learning and discovery in personalised recommendations. 
In this section, we propose strategies starting from purely personalised recommendations (baseline) to progressively introduce joint optimisation objectives. We work with a dataset described in \autoref{sec:dataset} that comprises several painting collections from the National Gallery, London.\footnote{\url{https://www.nationalgallery.org.uk/}}

\subsection{Personalised recommendations}

Let $\calP = \{p_1, p_2, \dots, p_m\}$ be a set of paintings with their associated embeddings (latent feature vectors)  $\calP^* = \{\vp_1, \vp_2, \dots, \vp_m\}$ according to CLIP or BLIP, and let $\calP^u  = \{p^u_1, p^u_2, \dots, p^u_n\}$ be the set of paintings a user $u$ has rated,  where $\calP^u \subset \calP$ and $\omega^u=\{\omega^u_1, \omega^u_2, \dots, \omega^u_n\}$
are the normalised ratings\footnote{In our work, users elicit their preferences in a 5-point scale rating (higher is better), thus we transform those values into weights $\omega^{u}_i \in [0,1]$ for every painting $p^{u}_i$ the user has rated.
} that $u$ gave to a small set of paintings $\calP^u$.
Once the dataset embeddings are learned using our backbone models (CLIP or BLIP), we compute the similarity matrix $\mathbf{A}$ for all the paintings as discussed in \autoref{sec:mmorl}. Then, the predicted score $S^u(p_i)$ the user would give to each painting in the collection $P$ is calculated based on the weighted average distance between the rated paintings and all other paintings:
\begin{equation}\label{eq:user-score}
    S^u(p_i) =  \frac{1}{n} \sum^{n}_{j = 1} \left( \omega^u_j \, \mathbf{A}_{ij} \right)
\end{equation}
where $\mathbf{A}_{ij} = d(\vp_i, \vp_j)$ is the similarity score between embeddings of paintings $p_i$ and $p_j$ in the computed similarity matrix. 
The summation in \autoref{eq:user-score} is taken over all user's rated paintings $n = |\calP^u|$.
Once the scoring procedure is complete, the paintings are sorted
and the $r$ most similar paintings constitute a ranked recommendation list (in this work we use $r=9$).
In sum, a VA RecSys task consists of recommending the most similar paintings to a user based on a small set of paintings rated before, i.e., the elicited preferences.

\subsection{Popular recommendations}
In addition to individual preferences, users also vary in their interest to visit famous paintings. 
To account for this, we introduce a popularity score, denoted by $S^\text{pop}(p_i)$, for each painting in our dataset. 
This score is based on a ``must-see'' list, generated according to public reviews from the National Gallery website. 
To consider both the user's individual preferences and the additional influence of the crowd on the popularity of a painting, 
we create an aggregated preference score $S^u_{+}(p_i)$ for each painting in the collection:
\begin{equation}
 S^u_{+}(p_i) =  S^u(p_i) + \beta \;  S^\text{pop}(p_i) \label{eq:pop}
\end{equation}
where $\beta$ is a user-provided hyperparameter, 
which determines their level of interest in (or their tolerance to) seeing popular items.
Note that $\beta$ is considered as a boosting scoring method,
since popularity has a direct influence on the user recommendations.

\subsection{Representative recommendations}
\label{subsec:fair}
As mentioned before, other key stakeholders are museum curators, responsible for preparing the art exhibitions,
therefore VA recommendations should reflect certain themes or topics that curators may want to promote.  
The National Gallery dataset, for example, features nine curated stories defined by artists (\textit{Women’s Lives, Contemporary Style \& Fashion, Water, Women Artists \& Famous Women, Warfare, Monsters \& Demons, Migration \& Exile, Death, Battles \& Commanders}). Each story constitutes a certain number of paintings from the collection. 
The objective of the curators here is to increase the number of curated stories in the recommendation set; 
i.e., the recommended set contains paintings that are representative or fairly selected from the curated story groups. Thus, we define a representative story selection strategy, adopted from Mehrotra et al.~\cite{mehrotra2015representative}.
A recommendation set $\calR$ is fairly representative 
if it contains paintings that belong to different story groups in a balanced way,
so a representative story selection function $\psi(\calR)$ is given by: 
\begin{equation}
 \psi(\calR)=  \sum ^{K}_{a = 1} \sqrt{\sum_{p_i \in \calS_a \cap \calR}^{}} \gamma_{p_i}
\end{equation}
where $K$ is the total number of story groups, $\calS_a$ is the $a$th story group and $\gamma_{p_i}$ is a count for every painting $p_i$ selected from a story group $i$. The function $\psi(\calR)$ rewards recommendation sets that are diverse in terms of the different story groups covered. 

It can be noticed that the representativeness score for a set that contains paintings belonging to only one or few of the story groups is lower than a set that covers all or most of the story groups, owing to the square root function. For example assuming three story groups ($K = 3$) a recommendation set $\calR$ that chooses 2 paintings from $\calS_1$ and 1 painting from each of $\calS_2$ and $\calS_3$ gets a higher $\psi(\calR)$ score 
as compared to a recommendation set that chooses 4 paintings from just $\calS_1$, 
since $\sqrt{2} + \sqrt{1} + \sqrt{1} > \sqrt{4} + \sqrt{0} + \sqrt{0})$.

\subsection{MOSAIC policies}

The discussion above yields three different policies worth of investigation, 
related to our key VA RecSys stakeholders. 
The first policy relates to matching user preferences.
Given a set of paintings $\calP$ and a user $u$, we select the
most relevant set $\calR$ to recommend, which maximises:  
\begin{equation}
\text{Policy}\;1: 
 \left\{\begin{matrix}
\arg\max  \sum_{i=1}^{\calR}  S^u(p_i) 
\end{matrix}\right.\label{pol1}
\end{equation}
The second policy comes from the crowd influence on the popularity of paintings, which maximises: 
\begin{equation}
\text{Policy}\;2: 
 \left\{\begin{matrix}
\arg\max  \sum_{i=1}^{\calR} S^\text{pop}(p_i) 
\end{matrix}\right.\label{pol2}
\end{equation}
The third policy investigates matching the curators' goal of finding a recommendation set 
that features representative paintings from each of the story groups, which maximises:
\begin{equation}
\text{Policy}\;3: 
 \left\{\begin{matrix}
\arg\max\  \psi(\calR)= \arg\max \sum ^{K}_{a = 1} \sqrt{\sum_{p_i \in \calS_a \cap \calR}^{}}\gamma _{p_i}
\end{matrix}\right.\label{pol3}
\end{equation}
Note that maximising over $\psi(\calR)$ ensures diversity of the recommendations in terms of the story groups to be covered. However, minimising over $\psi(\calR)$ ensures consistency of the recommendations in terms of multimodal information.
Also note that contrary to the influence of the crowd,\footnote{Making items popular is not an objective of the crowd, but rather a consequence of their behaviour, indirectly influencing user preferences.}  curators and users might have competing objectives, therefore our formulation accounts for that by introducing a complementary dependency through a $\xi$ parameter,
as described in the next section.

\subsection{MOSAIC engines}

With the goal of introducing multistakeholder awareness in VA RecSys, we designed different versions of MOSAIC engines by gradually departing from optimizing for user preferences and combining the above policies. We start by introducing the impact of popularity on personalised recommendations. Thus, we combine Policies 1 and 2
and create two VA RecSys engines: CLIP-P and BLIP-P.
These engines use CLIP and BLIP respectively as a backbone multimodal engines to compute the user's preference scores $S^u(p_i)$ according to the scoring function in \autoref{eq:pop}.

To study the impact of representativeness on personalised recommendations, we combine Policies 1 and 3 
and create two new VA RecSys engines: CLIP-F and BLIP-F.
These engines try jointly to satisfy individual user preferences and curator goals. 
This translates to finding a recommendation set $\calR$ of paintings by solving the following Mixed Integer Programming (MIP) problem:
\begin{equation}
\text{MIP}_{1} = \arg\max \left (   
  (1 - \xi)  
\sum_{i=1}^{R}  p_i \, S^u(p_i)
+ \xi \sum ^{K}_{a = 1} \sqrt{\sum_{p_i \in S_a \cap \calR}^{}}\gamma _{p_i}  \right )\label{eq:fair}
\end{equation}
where $\xi$ is a user-provided hyperparameter, indicating their tolerance to receiving diverse painting recommendations.

Finally, by combining all the three policies above, we introduce two full-versions of MOSAIC, CLIP-M and BLIP-M,
each using either one of the backbone architectures as preference scoring models. 
Ultimately, MOSAIC aims for a recommended set $\calR$ that concurrently satisfies the above three policies (i.e., individual user preferences, crowd influence in terms of popularity bias, and curators' objective in terms of maximising the number of story groups represented in the recommendation set) achieved by solving the following MIP problem:
\begin{equation}
\text{MIP}_{2} = \arg\max \left (   
  (1 - \xi) 
\sum_{i=1}^{\calR} p_i \, S^u_{+}(p_i)
+ \xi \sum ^{K}_{a = 1} \sqrt{\sum_{p_i \in \calS_a \cap \calR}^{}}\gamma _{p_i}  \right )\label{eq:mosaic}
\end{equation}
where $S^u_{+}(p_i)$ combines Policy 1 and 2 
and $\xi$ indicates the user's tolerance to receiving diverse painting recommendations. 

Overall, we developed eight VA RecSys engines.
The first two engines, CLIP and BLIP, are considered baselines, 
since they generate pure personalised recommendations. 
Next, CLIP-P and BLIP-P factor in the influence of crowds, in terms of popularity bias. 
Then, CLIP-F and BLIP-F take into account the curator's preferences, in terms of fair (i.e., representative) content selection. 
Finally, the full-version of MOSAIC (CLIP-M and BLIP-M) combine the three key stakeholders considered relevant in this context, informed by Cultural Heritage studies \cite{kontiza2018museum, vlachidis2018crosscult, zhitomirsky2023they}: individual users, crowd of visitors, and museum curators.
Our proposed VA RecSys engines are outlined in Algorithms 1 to 5, respectively.

\begin{algorithm*}[!ht]
\caption{Dataset preprocessing.}
\begin{algorithmic}[1]
\small
    \Procedure{Preprocess}{$\text{Model}, P$} \Comment{Model $\in \{\text{CLIP, BLIP}\}$}
        \State $\calP \gets \Call{FeaturizePaintings}{\text{Model}, P}$ \Comment{Compute painting embeddings}
        \State $\mathbf{A} \gets \emptyset$ \Comment{Initialize $\mathbf{A} \in \mathbb{R}^{m \times m}$ where $m = |\calP|$}
        \For{$\vp_i$ and $\vp_j \in \calP$} \Comment{Iterate over collection}
            \State $\mathbf{A}_{ij} \gets$ \Call{Similarity}{$\vp_i, \vp_j$} \Comment{Compute embeddings similarity}
        \EndFor
        \State \Return $\mathbf{A}$
    \EndProcedure
\end{algorithmic}
\end{algorithm*}

\begin{algorithm*}[!ht]
\caption{
    Personalised VA RecSys (baseline engine).
}
\begin{algorithmic}[1]
\small
    \Require Similarity matrix $\mathbf{A}$ of featurised paintings
    \Procedure{RecommendPaintings}{$\text{Model}, P^u, \omega^u, r$} \Comment{Model $\in \{\text{CLIP, BLIP}\}$}
        \State $\calP^u \gets \Call{FeaturizePaintings}{\text{Model}, P^u}$ \Comment{Load computed embeddings, as $P^u \subset P$}
        \State $S^u \gets \emptyset$ \Comment{Initialize recommendations list}
        \For{$\vp_i \in \calP^u$ and $\vp_j \in \calP$} \Comment{Iterate over collection}
            \State $S^u(\vp_i) = \frac{1}{n} \sum^{n}_{j = 1} \omega^u_j \mathbf{A}_{ij}$ \Comment{Bias similarity matrix towards user's preferences}
        \EndFor
        \State \Call{Sort}{$S^u$} \Comment{Descending sort, as higher score means higher similarity}
        \State \Return \Call{Slice}{$S^u, r$} \Comment{Pick the first $r$ elements in the ranking}
    \EndProcedure
\end{algorithmic}
\end{algorithm*}

\begin{algorithm*}[!ht]
\caption{
    VA RecSys with popular recommendations (CLIP-P \& BLIP-P).
}
\begin{algorithmic}[1]
\small
    \Require Similarity matrix $\mathbf{A}$ of featurised paintings; Popularity score $S^\text{pop} \, \forall p_i \in P$
    \Procedure{RecommendPaintings}{$\text{Model}, P^u, \omega^u, \beta,  r$} \Comment{Model $\in \{\text{CLIP, BLIP}\}$}
        \State $\calP^u \gets \Call{FeaturizePaintings}{\text{Model}, P^u}$ \Comment{Load computed embeddings, as $P^u \subset P$}
        \State $S^u_{AG} \gets \emptyset$ \Comment{Initialize recommendations list}
        \For{$\vp_i \in \calP^u$ and $\vp_j \in \calP$} \Comment{Iterate over collection}
            \State $S^u(\vp_i) = \frac{1}{n} \sum^{n}_{j = 1} \omega^u_j \mathbf{A}_{ij}$ \Comment{Bias similarity matrix towards user's preferences}
            \State $S^u_{+}(\vp_i) =  S^u(\vp_i) + \beta \;  S^\text{pop}(\vp_i)$ \Comment{Compute aggregated preferences score}
        \EndFor
        \State \Call{Sort}{$S^u_{+}$} \Comment{Descending sort, as higher score means higher relevance}
        \State \Return \Call{Slice}{$S^u_{+}, r$} \Comment{Pick the first $r$ elements in the ranking}
    \EndProcedure
\end{algorithmic}
\end{algorithm*}

\begin{algorithm*}[!ht]
\caption{VA RecSys with representative recommendations (CLIP-F \& BLIP-F).}
\begin{algorithmic}[1]
\small
    \Require Similarity matrix $\mathbf{A}$ of featurised paintings
    \Procedure{RecommendPaintings}{$\text{Model}, P^u, \omega^u, \xi, \gamma_{p_i}, K, r$}
        \Comment{Model $\in \{\text{CLIP, BLIP}\}$; $\forall p_i \in P$}
        \State $\calP^u \gets \Call{FeaturizePaintings}{\text{Model}, P^u}$ \Comment{Load computed embeddings, as $P^u \subset P$}
        \State $S^u \gets \emptyset$ \Comment{Initialize recommendations list}
        \For{$\vp_i \in \calP^u$ and $\vp_j \in \calP$} \Comment{Iterate over collection}
            \State $S^u(\vp_i) = \frac{1}{n} \sum^{n}_{j = 1} \omega^u_j \mathbf{A}_{ij}$ \Comment{Bias similarity matrix towards user's preferences}
        \EndFor
        \State $\calR \gets$ \Call{Solve $\text{MIP}_1$}{$\text{Model},S^u,  P^u, \omega^u, \xi,  \gamma_{p_i}, K, r$} \Comment{Generate optimal recommendation set $\calR = \{p_1, \dots, p_r\}$}
        \State \Return {$\calR$} %\Comment{Pick the first $r$ elements in the fused ranking}
    \EndProcedure
\end{algorithmic}
\end{algorithm*}

\begin{algorithm*}[!ht]
\caption{VA RecSys with popular and representative recommendations (CLIP-M \& BLIP-M).}
\begin{algorithmic}[1]
\small
    \Require Similarity matrix $\mathbf{A}$ of featurised paintings; Popularity score $S^\text{pop} \, \forall p_i \in P$
    \Procedure{RecommendPaintings}{$\text{Model}, P^u, \omega^u, \beta,  \xi, \gamma_{p_i}, K, r$}
        \Comment{Model $\in \{\text{CLIP, BLIP}\}$; $\forall p_i \in P$}
        \State $\calP^u \gets \Call{FeaturizePaintings}{\text{Model}, P^u}$ \Comment{Load computed embeddings, as $P^u \subset P$}
        \State $S^u_{+} \gets \emptyset$ \Comment{Initialize recommendations list}
        \For{$\vp_i \in \calP^u$ and $\vp_j \in \calP$} \Comment{Iterate over collection}
            \State $S^u(p_i) = \frac{1}{n} \sum^{n}_{j = 1} \omega^u_j \mathbf{A}_{ij}$ \Comment{Bias similarity matrix towards user's preferences}
            \State $S^u_{+}(p_i) =  S^u(p_i) + \beta \;  S^\text{pop}(p_i)$ \Comment{Compute aggregate preferences score}
        \EndFor
        \State $\calR \gets$ \Call{Solve $\text{MIP}_2$}{$\text{Model},S^u_{+},  P^u, \omega^u, \xi, \gamma_{p_i}, K, r$} \Comment{Generate optimal recommendation set $\calR = \{p_1, \dots, p_r\}$}  
        \State \Return {$\calR$} %\Comment{Pick the first $r$ elements in the fused ranking}
    \EndProcedure
\end{algorithmic}
\end{algorithm*}

\subsection{Modality Gap}
Multimodal models map inputs from different data modalities (e.g. image and text) into a shared latent representation space~\cite{li2022blip, radford2021learning}. Although multi-modal approaches like CLIP and BLIP have demonstrated their effectiveness on various tasks, a recent study by Liang et al.~\cite{NEURIPS2022_702f4db7} has shown the presence of a modality gap on their representation spaces, where embeddings from different modalities are located in completely separate regions of the embedding space. Liang et al. also demonstrated that a variation in this gap is caused by a combination of model initialisation and contrastive learning optimisation, significantly affecting performance in downstream tasks.  

This modality gap effect has been shown to be more prevalent on models using only contrasting learning like CLIP as compared to BLIP,  which optimises not only contrastive loss (ITC) but also other objectives jointly (ITM and LM)~\cite{NEURIPS2022_702f4db7}. Hence, this provides supporting evidence for the fact that BLIP outperforms CLIP on various downstream tasks~\cite{shi2023upop}, including VA RecSys~\cite{YilmaleivaUMAP23}.

\begin{postulate}
\label{post:one}
\textit{We know that a reduced modality gap has a positive impact on downstream performance~\cite{NEURIPS2022_702f4db7} and that BLIP outperforms CLIP in VA RecSys tasks~\cite{YilmaleivaUMAP23}. Therefore we expect all BLIP-based engines to achieve higher performance than the ones using CLIP.}
\end{postulate}

\begin{postulate}
\label{post:two}
\textit{We expect Postulate \autoref{post:one} to be true unless the introduction of popularity and representativeness compensate for the modality gap effect on VA RecSys tasks. }
\end{postulate}

%% =============================================================================
\section{Offline evaluation}
\label{sec:eval1}
%% =============================================================================

We conducted a series of experiments to assess the effectiveness of the three proposed MOSAIC variants 
when it comes to seamlessly blending learning and discovery with personal user preferences. 
Concretely, we measured the impact of introducing popularity 
and representativeness (fair story selection) strategies in both baseline VA RecSys engines CLIP and BLIP.

All algorithms were implemented using Python 3.9 with Gurobi 10.0\footnote{\url{https://www.gurobi.com/}} 
to solve the core MIP problems. 
All experiments were run on a 3.10\,GHz Intel(R) Xeon(R) Gold 6254 CPU with 4 cores and 16\,GB of RAM. 

\subsection{Dataset}
\label{sec:dataset}

\begin{figure*}[!ht]
\centering
\includegraphics[width= \textwidth]{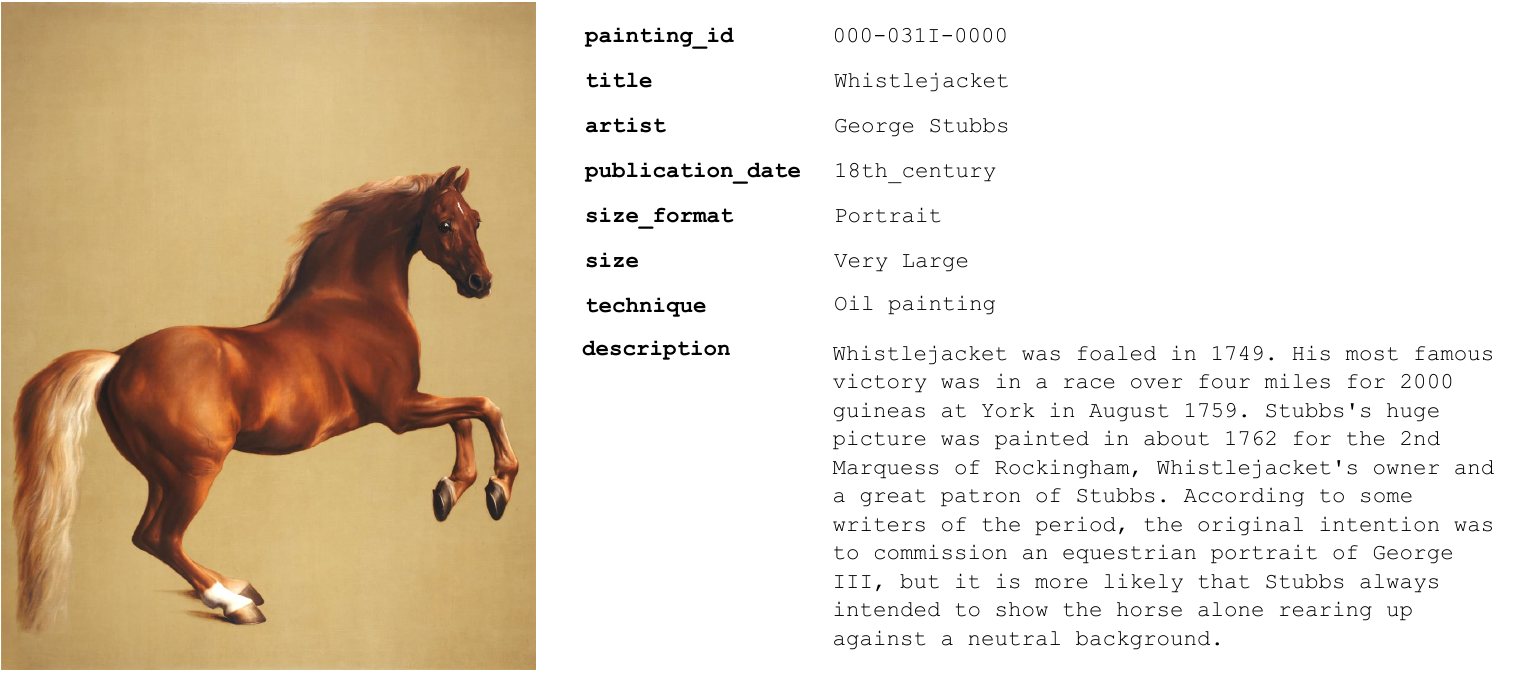}
\caption{Sample painting and associated metadata from the National Gallery dataset.}
\label{fig:ng_sample}
\end{figure*}

We used a public dataset containing 2,368 paintings from The National Gallery, London 
to train our backbone models BLIP and CLIP. 
This curated set of paintings belongs to the CrossCult Knowledge Base.\footnote{\url{https://www.crosscult.lu/}} 
Each painting image is accompanied by a set of text-based metadata, 
which makes this dataset suitable for multimodal feature learning approaches. 
A sample data point is shown in \autoref{fig:ng_sample}. 
For learning textual information, we considered all available painting attributes
such as artist name, painting title, technique used, etc. combined with the description provided by museum curators. 
The dataset also provides nine story groups, as previously mentioned in \autoref{subsec:fair},
and 213 elicited user preferences, collected in previous studies \cite{Yilmaleiva23, YilmaleivaUMAP23} making it the most suitable dataset for this study as other publicly available art datasets such as Smithsonian\footnote{\url{https://library.si.edu/data}} lack such user preference information.

\subsection{Procedure}

We started by processing the aforementioned elicited user preferences in the dataset.
Users provided 5-star ratings for up to nine paintings each (one painting per story group), 
that we normalized in the [0,1] range.
Next, since the dataset has no user preferences related to popularity and representativeness, we simulated them.
Specifically, from the users' perspective, 
popularity is defined as the tolerance (or willingness) to receiving popular paintings,
whereas representativeness is defined as the tolerance to receiving diverse paintings,
that are fairly selected from the curated story groups.
Then, tolerance to both popular and diverse (representative selection) of paintings 
is simulated by sampling from a Poisson distribution. We simulated three scenarios, using mean rates of $0$, $0.5$, and $1$,
where 0 denotes no tolerance and 1 denotes maximal tolerance. 

These simulated tolerances are the values we use for $\beta$ and $\xi$ in the joint optimisation objectives, 
as a user-provided hyperparameter discussed in \autoref{sec:method}. 
This way, we can anticipate the performance of MOSAIC under different use case scenarios.
For example, $\beta = 1$ means that the user is highly interested in receiving popular painting recommendations. 
Thus, more weight is given to the second half of the objective function in \autoref{eq:pop} 
to promote the recommendation of popular paintings. 
Similarly, $\xi = 0$  means that a user is not interested in receiving diverse recommendations, 
thus the first half of the objective function in \autoref{eq:fair} and \autoref{eq:mosaic} 
will be optimised to promote the consistency of recommendations with user preferences 
instead of promoting a representative selection of paintings.

\subsection{Results}
\label{subsec: offline_results}

We compute the Jaccard index (Intersection over Union, IoU) and Rank-Biased Overlap (RBO) 
in a pairwise fashion between the generated recommendations from each MOSAIC variant. 
Both IoU and RBO are widely used measures in information retrieval~\cite{10.1145/1852102.1852106} as they complement each other: IoU measures differences between sets, whereas RBO measures differences between rankings.\footnote{The recommendations are assessed holistically, therefore IoU is an informative measure to our study.}However, the recommendations are actually ranked, therefore RBO is also an informative measure to report.
This assessment of MOSAIC is particularly relevant to see the influence of introducing popularity (policy 2) and representative selection (policy 3) on personalised recommendations (policy 1).
Tables \ref{tbl:res-clip} and \ref{tbl:res-blip} summarise the results reporting the mean values of IoU and RBO along with their standard deviation (SD), which gives an overview about the dispersion of the measures.

\begin{sidewaystable}[htbp]
\centering
\large
\begin{minipage}{\textwidth}
\resizebox{\columnwidth}{!}{%
\begin{tabular}{lcccccccccccc}
\cmidrule(l){2-13}
 &
  \multicolumn{1}{l}{\system{CLIP}} &
  \multicolumn{1}{l}{\systempop{CLIP}{0}} &
  \multicolumn{1}{l}{\systempop{CLIP}{0.5}} &
  \multicolumn{1}{l}{\systempop{CLIP}{1}} &
  \multicolumn{1}{l}{\systemfair{CLIP}{0}} &
  \multicolumn{1}{l}{\systemfair{CLIP}{0.5}} &
  \multicolumn{1}{l}{\systemfair{CLIP}{1}} &
  \multicolumn{1}{l}{\mosaic{CLIP}{0}{0}} &
  \multicolumn{1}{l}{\mosaic{CLIP}{0.5}{0.5}} &
  \multicolumn{1}{l}{\mosaic{CLIP}{1}{1}} &
  \multicolumn{1}{l}{\mosaic{CLIP}{0}{1}} &
  \multicolumn{1}{l}{\mosaic{CLIP}{1}{0}} \\ \cmidrule(l){2-13} 
\system{CLIP}             &             & \colorbox{pink}{0.01 $\pm$ 0.38} & \colorbox{pink}{0.01 $\pm$ 0.08}        &\colorbox{pink} {0.01 $\pm$ 0.38} &  \colorbox{pink}{0.63 $\pm$ 0.28}  & \colorbox{pink}{0.85 $\pm$ 0.24 }       &\colorbox{pink}{0.97 $\pm$ 0.12} & \colorbox{pink}{0.65 $\pm$ 0.25} & \colorbox{pink}{0.87 $\pm$ 0.20} & \colorbox{pink}{0.95 $\pm$ 0.12 } & \colorbox{pink}{0.95 $\pm$ 0.12} & \colorbox{pink}{0.64 $\pm$ 0.28} \\
\systempop{CLIP}{0}       & \colorbox{cyan}{0.01 $\pm$ 0.02} &   &  \colorbox{pink}{0.99 $\pm$ 0.07}&\colorbox{pink}{1.00  $\pm$ 0.00} & \colorbox{pink}{0.01 $\pm$ 0.03} & \colorbox{pink}{0.01 $\pm$ 0.03 }    &\colorbox{pink}{0.01 $\pm$ 0.03} & \colorbox{pink}{0.01 $\pm$ 0.03} & \colorbox{pink}{0.01 $\pm$ 0.04} & \colorbox{pink}{0.01 $\pm$ 0.03} & \colorbox{pink}{0.01 $\pm$ 0.03} & \colorbox{pink}{0.01 $\pm$ 0.03} \\
\systempop{CLIP}{0.5}       & \colorbox{cyan}{0.01 $\pm$ 0.02} & \colorbox{cyan}{0.99 $\pm$ 0.07} &        & \colorbox{pink}{0.99 $\pm$ 0.06} & \colorbox{pink}{0.01  $\pm$ 0.06}    & \colorbox{pink}{ 0.01 $\pm$ 0.05}  &\colorbox{pink}{0.01  $\pm$ 0.08} & \colorbox{pink}{0.01  $\pm$ 0.03} & \colorbox{pink}{1.00 $\pm$ 0.06} & \colorbox{pink}{0.01  $\pm$ 0.08} & \colorbox{pink}{0.01  $\pm$ 0.06} & \colorbox{pink}{0.01 $\pm$ 0.05} \\
\systempop{CLIP}{1}       & \colorbox{cyan}{0.01  $\pm$ 0.04} & \colorbox{cyan}{1.00  $\pm$ 0.00} &  \colorbox{cyan}{0.99  $\pm$ 0.06}&         & \colorbox{pink}{0.01  $\pm$ 0.03}  & \colorbox{pink}{ 0.01 $\pm$ 0.03}  &\colorbox{pink}{0.01  $\pm$ 0.04} & \colorbox{pink}{0.01  $\pm$ 0.04} & \colorbox{pink}{0.01  $\pm$ 0.04} & \colorbox{pink}{0.01  $\pm$ 0.04}  & \colorbox{pink}{0.01  $\pm$ 0.04} & \colorbox{pink}{0.01  $\pm$ 0.04} \\
\systemfair{CLIP}{0}      & \colorbox{cyan}{0.51  $\pm$ 0.29} & \colorbox{cyan}{0.01  $\pm$ 0.02} &  \colorbox{cyan}{0.01  $\pm$ 0.05} & \colorbox{cyan}{0.04  $\pm$ 0.02} &       & \colorbox{pink}{ 0.67 $\pm$ 0.24}        &\colorbox{pink}{0.72  $\pm$ 0.25} & \colorbox{pink}{0.82  $\pm$ 0.21} & \colorbox{pink}{0.71  $\pm$ 0.26} & \colorbox{pink}{0.66  $\pm$ 0.28}  & \colorbox{pink}{0.66  $\pm$ 0.27} & \colorbox{pink}{0.82  $\pm$ 0.20} \\
\systemfair{CLIP}{0.5}      & \colorbox{cyan}{0.79 $\pm$ 0.30} & \colorbox{cyan}{0.01 $\pm$ 0.02} &\colorbox{cyan}{0.01 $\pm$ 0.07}&\colorbox{cyan}{0.04  $\pm$ 0.02} & \colorbox{cyan}{0.52  $\pm$ 0.32}         &        &\colorbox{pink}{0.94  $\pm$ 0.12} & \colorbox{pink}{0.54  $\pm$ 0.23} & \colorbox{pink}{0.94  $\pm$ 0.12} & \colorbox{pink}{1.00  $\pm$ 0.00}  & \colorbox{pink}{0.94  $\pm$ 0.12} & \colorbox{pink}{0.54  $\pm$ 0.23} \\
\systemfair{CLIP}{1}      & \colorbox{cyan}{0.95  $\pm$ 0.15} &\colorbox{cyan}{0.01 $\pm$ 0.02} & \colorbox{cyan}{0.01  $\pm$ 0.06} & \colorbox{cyan}{0.01  $\pm$ 0.02} & \colorbox{cyan}{0.60  $\pm$ 0.30} & \colorbox{cyan}{0.91  $\pm$ 0.17}   &            & \colorbox{pink}{0.85  $\pm$ 0.24} & \colorbox{pink}{0.69  $\pm$ 0.26} & \colorbox{pink}{0.80  $\pm$ 0.24}  & \colorbox{pink}{0.85  $\pm$ 0.23} & \colorbox{pink}{0.69  $\pm$ 0.26} \\
\mosaic{CLIP}{0}{0} & \colorbox{cyan}{0.52  $\pm$ 0.30} & \colorbox{cyan}{0.01 $\pm$ 0.02} &\colorbox{cyan}{0.01  $\pm$ 0.03}& \colorbox{cyan}{0.01  $\pm$ 0.02} & \colorbox{cyan}{0.73  $\pm$ 0.28}  & \colorbox{cyan}{0.79  $\pm$ 0.30}        &\colorbox{cyan}{0.39  $\pm$ 0.29} &             & \colorbox{pink}{0.72  $\pm$ 0.25} & \colorbox{pink}{0.68  $\pm$ 0.28}  & \colorbox{pink}{0.68  $\pm$ 0.25} & \colorbox{pink}{0.83  $\pm$ 0.20} \\
\mosaic{CLIP}{0.5}{0.5} & \colorbox{cyan}{0.80  $\pm$ 0.28} & \colorbox{cyan}{0.01 $\pm$ 0.03} &\colorbox{cyan}{0.01  $\pm$ 0.08}&\colorbox{cyan}{0.01  $\pm$ 0.02} & \colorbox{cyan}{0.59  $\pm$ 0.31}  & \colorbox{cyan}{0.58  $\pm$ 0.29}      & \colorbox{cyan}{0.60  $\pm$ 0.30} & \colorbox{cyan}{0.40  $\pm$ 0.29} &             & \colorbox{pink}{0.86  $\pm$ 0.19}  & \colorbox{pink}{0.85  $\pm$ 0.20} & \colorbox{pink}{0.71  $\pm$ 0.26} \\
\mosaic{CLIP}{1}{1} & \colorbox{cyan}{0.93  $\pm$ 0.17} & \colorbox{cyan}{0.01 $\pm$ 0.02} &\colorbox{cyan}{0.01  $\pm$ 0.04}&\colorbox{cyan}{0.01  $\pm$ 0.02} & \colorbox{cyan}{0.53  $\pm$ 0.31}  & \colorbox{cyan}{0.72  $\pm$ 0.31}      & \colorbox{cyan}{0.56  $\pm$ 0.31} & \colorbox{cyan}{0.39  $\pm$ 0.29} & \colorbox{cyan}{0.80  $\pm$ 0.27} &              & \colorbox{pink}{0.92  $\pm$ 0.16} & \colorbox{pink}{0.66  $\pm$ 0.27} \\
\mosaic{CLIP}{0}{1} & \colorbox{cyan}{0.95  $\pm$ 0.14} & \colorbox{cyan}{0.01 $\pm$ 0.02} &\colorbox{cyan}{0.01  $\pm$ 0.06}&\colorbox{cyan}{0.01  $\pm$ 0.02} & \colorbox{cyan}{0.54  $\pm$ 0.32}  & \colorbox{cyan}{0.77  $\pm$ 0.29}       & \colorbox{cyan}{0.56  $\pm$ 0.31} & \colorbox{cyan}{0.57  $\pm$ 0.31} & \colorbox{cyan}{0.89  $\pm$ 0.23} & \colorbox{cyan}{1.00  $\pm$ 0.00}  &             & \colorbox{pink}{0.66  $\pm$ 0.27} \\
\mosaic{CLIP}{1}{0} & \colorbox{cyan}{0.52  $\pm$ 0.31} & \colorbox{cyan}{0.01 $\pm$ 0.03} &\colorbox{cyan}{0.01  $\pm$ 0.08} & \colorbox{cyan}{0.01  $\pm$ 0.02} & \colorbox{cyan}{0.73  $\pm$ 0.27}  & \colorbox{cyan}{0.58  $\pm$ 0.31}        &\colorbox{cyan}{0.74  $\pm$ 0.26} & \colorbox{cyan}{0.74  $\pm$ 0.26} & \colorbox{cyan}{0.59  $\pm$ 0.31} & \colorbox{cyan}{0.54  $\pm$ 0.32}  & \colorbox{cyan}{0.54  $\pm$ 0.31} &  \\ 
\bottomrule
\end{tabular}
}%
\caption{Offline evaluation of CLIP-based MOSAIC engines. 
    We report Mean $\pm$ SD of the following pairwise ranking similarity measures:
    Jaccard index (IoU) \raisebox{.25em}{\colorbox{cyan}{ }}
    and Rank-biased overlap (RBO) \raisebox{.25em}{\colorbox{pink}{ }}.
}
\label{tbl:res-clip}
\vspace{1cm}
\end{minipage}
\hfill
\begin{minipage}{\textheight}
\centering
\large
\resizebox{\columnwidth}{!}{%
\begin{tabular}{lcccccccccccc}
\cmidrule(l){2-13}
 &
  \multicolumn{1}{l}{\system{BLIP}} &
  \multicolumn{1}{l}{\systempop{BLIP}{0}} &
  \multicolumn{1}{l}{\systempop{BLIP}{0.5}} &
  \multicolumn{1}{l}{\systempop{BLIP}{1}} &
  \multicolumn{1}{l}{\systemfair{BLIP}{0}} &
  \multicolumn{1}{l}{\systemfair{BLIP}{0.5}} &
  \multicolumn{1}{l}{\systemfair{BLIP}{1}} &
  \multicolumn{1}{l}{\mosaic{BLIP}{0}{0}} &
  \multicolumn{1}{l}{\mosaic{BLIP}{0.5}{0.5}} &
  \multicolumn{1}{l}{\mosaic{BLIP}{1}{1}} &
  \multicolumn{1}{l}{\mosaic{BLIP}{0}{1}} &
  \multicolumn{1}{l}{\mosaic{BLIP}{1}{0}} \\ \cmidrule(l){2-13}
\system{BLIP}               &             & \colorbox{pink}{0.16 $\pm$ 0.32} & \colorbox{pink}{0.04 $\pm$ 0.16}       &\colorbox{pink}{0.03 $\pm$ 0.13} & \colorbox{pink}{0.99 $\pm$ 0.03} & \colorbox{pink}{0.99 $\pm$ 0.03} & \colorbox{pink}{0.99 $\pm$ 0.03} & \colorbox{pink}{0.99 $\pm$ 0.03} & \colorbox{pink}{0.99 $\pm$ 0.03} & \colorbox{pink}{0.99 $\pm$ 0.03}  & \colorbox{pink}{0.99 $\pm$ 0.05} & \colorbox{pink}{0.99 $\pm$ 0.03} \\
\systempop{BLIP}{0}       & \colorbox{cyan}{0.16 $\pm$ 0.30} &               &  \colorbox{pink}{0.83 $\pm$ 0.32}       & \colorbox{pink}{0.84 $\pm$ 0.33} & \colorbox{pink}{0.16 $\pm$ 0.32}    & \colorbox{pink}{0.16 $\pm$ 0.32}          &\colorbox{pink}{0.16 $\pm$ 0.32} & \colorbox{pink}{0.16 $\pm$ 0.32} & \colorbox{pink}{0.16 $\pm$ 0.32} & \colorbox{pink}{0.16 $\pm$ 0.32} & \colorbox{pink}{0.16 $\pm$ 0.32} & \colorbox{pink}{0.16 $\pm$ 0.30} \\
\systempop{BLIP}{0.5}       & \colorbox{cyan}{0.04 $\pm$ 0.15} & \colorbox{cyan}{0.74 $\pm$ 0.37} &        & \colorbox{pink}{0.94 $\pm$ 0.18} & \colorbox{pink}{0.04 $\pm$ 0.14}    & \colorbox{pink}{0.04 $\pm$ 0.14}  &\colorbox{pink}{0.04 $\pm$ 0.14} & \colorbox{pink}{0.04 $\pm$ 0.14} & \colorbox{pink}{0.04 $\pm$ 0.14} & \colorbox{pink}{0.04 $\pm$ 0.14} & \colorbox{pink}{0.04 $\pm$ 0.14} & \colorbox{pink}{0.04 $\pm$ 0.14} \\
\systempop{BLIP}{1}       & \colorbox{cyan}{0.02 $\pm$ 0.12} & \colorbox{cyan}{0.79 $\pm$ 0.36} & \colorbox{cyan}{0.91 $\pm$ 0.23} &           & \colorbox{pink}{0.03 $\pm$ 0.13} &  \colorbox{pink}{0.03 $\pm$ 0.13} & \colorbox{pink}{0.03 $\pm$ 0.13} & \colorbox{pink}{0.03 $\pm$ 0.13} & \colorbox{pink}{0.03 $\pm$ 0.12} & \colorbox{pink}{0.03 $\pm$ 0.13}  & \colorbox{pink}{0.03 $\pm$ 0.13} & \colorbox{pink}{0.03 $\pm$ 0.12} \\
\systemfair{BLIP}{0}      & \colorbox{cyan}{0.99 $\pm$ 0.03} & \colorbox{cyan}{0.16 $\pm$ 0.30} & \colorbox{cyan}{0.04 $\pm$ 0.15}        & \colorbox{cyan}{0.03 $\pm$ 0.12} &             &  \colorbox{pink}{0.99 $\pm$ 0.00} &\colorbox{pink}{0.99 $\pm$ 0.01} & \colorbox{pink}{1.00 $\pm$ 0.00} & \colorbox{pink}{0.99 $\pm$ 0.07} & \colorbox{pink}{0.99 $\pm$ 0.07}  & \colorbox{pink}{0.99 $\pm$ 0.07} & \colorbox{pink}{1.00 $\pm$ 0.00} \\
\systemfair{BLIP}{0.5}      & \colorbox{cyan}{0.99 $\pm$ 0.03} & \colorbox{cyan}{0.16 $\pm$ 0.30} & \colorbox{cyan}{0.04 $\pm$ 0.15}        & \colorbox{cyan}{0.03 $\pm$ 0.12} & \colorbox{cyan}{0.99 $\pm$ 0.02}            &           &\colorbox{pink}{0.99 $\pm$ 0.01} & \colorbox{pink}{0.99 $\pm$ 0.01} & \colorbox{pink}{0.98 $\pm$ 0.00} & \colorbox{pink}{0.99 $\pm$ 0.01}  & \colorbox{pink}{0.99 $\pm$ 0.05} & \colorbox{pink}{0.99 $\pm$ 0.01} \\
\systemfair{BLIP}{1}      & \colorbox{cyan}{0.99 $\pm$ 0.03} & \colorbox{cyan}{0.16 $\pm$ 0.30} & \colorbox{cyan}{0.04 $\pm$ 0.15}         &\colorbox{cyan}{0.03 $\pm$ 0.12} & \colorbox{cyan}{0.99 $\pm$ 0.04} & \colorbox{cyan}{0.98 $\pm$ 0.05}          &            & \colorbox{pink}{0.99 $\pm$ 0.01} & \colorbox{pink}{0.99 $\pm$ 0.01} & \colorbox{pink}{0.99 $\pm$ 0.01}  & \colorbox{pink}{0.99 $\pm$ 0.03} & \colorbox{pink}{0.99 $\pm$ 0.01} \\
\mosaic{BLIP}{0}{0} & \colorbox{cyan}{0.99 $\pm$ 0.03} & \colorbox{cyan}{0.16 $\pm$ 0.30} & \colorbox{cyan}{0.04 $\pm$ 0.15}         & \colorbox{cyan}{0.03 $\pm$ 0.12} & \colorbox{cyan}{1.00 $\pm$ 0.00} & \colorbox{cyan}{0.99 $\pm$ 0.02}          &\colorbox{cyan}{0.99 $\pm$ 0.04} &             & \colorbox{pink}{0.99 $\pm$ 0.00} & \colorbox{pink}{0.99 $\pm$ 0.00}  & \colorbox{pink}{0.99 $\pm$ 0.03} & \colorbox{pink}{1.00 $\pm$ 0.00} \\
\mosaic{BLIP}{0.5}{0.5} & \colorbox{cyan}{0.99 $\pm$ 0.03} & \colorbox{cyan}{0.16 $\pm$ 0.30} & \colorbox{cyan}{0.04 $\pm$ 0.15}         & \colorbox{cyan}{0.03 $\pm$ 0.12} & \colorbox{cyan}{0.99 $\pm$ 0.07} &  \colorbox{cyan}{0.99 $\pm$ 0.04}         &\colorbox{cyan}{0.99 $\pm$ 0.04} & \colorbox{cyan}{0.99 $\pm$ 0.01} &             & \colorbox{pink}{0.99 $\pm$ 0.00}  & \colorbox{pink}{0.99 $\pm$ 0.03} & \colorbox{pink}{0.99 $\pm$ 0.00} \\
\mosaic{BLIP}{1}{1} & \colorbox{cyan}{0.99 $\pm$ 0.03} & \colorbox{cyan}{0.16 $\pm$ 0.30} & \colorbox{cyan}{0.04 $\pm$ 0.15}         & \colorbox{cyan}{0.03 $\pm$ 0.12} & \colorbox{cyan}{0.99 $\pm$ 0.07} &  \colorbox{cyan}{0.99 $\pm$ 0.05}         &\colorbox{cyan}{0.98 $\pm$ 0.03} & \colorbox{cyan}{0.99 $\pm$ 0.07} & \colorbox{cyan}{0.99 $\pm$ 0.03} &              & \colorbox{pink}{0.99 $\pm$ 0.03} & \colorbox{pink}{0.99 $\pm$ 0.00} \\
\mosaic{BLIP}{0}{1} & \colorbox{cyan}{0.98 $\pm$ 0.08} & \colorbox{cyan}{0.16 $\pm$ 0.30} &  \colorbox{cyan}{0.04 $\pm$ 0.15}         &\colorbox{cyan}{0.03 $\pm$ 0.12} & \colorbox{cyan}{0.99 $\pm$ 0.07} &   \colorbox{cyan}{0.99 $\pm$ 0.05}        &\colorbox{cyan}{0.98 $\pm$ 0.08} & \colorbox{cyan}{0.99 $\pm$ 0.07} & \colorbox{cyan}{0.99 $\pm$ 0.08} & \colorbox{cyan}{0.98 $\pm$ 0.08}  &             & \colorbox{pink}{0.99 $\pm$ 0.03} \\
\mosaic{BLIP}{1}{0} & \colorbox{cyan}{0.99 $\pm$ 0.02} & \colorbox{cyan}{0.16 $\pm$ 0.30} & \colorbox{cyan}{0.04 $\pm$ 0.15}          &\colorbox{cyan}{0.03 $\pm$ 0.12} & \colorbox{cyan}{1.00 $\pm$ 0.00} & \colorbox{cyan}{0.99 $\pm$ 0.05}          &\colorbox{cyan}{0.99 $\pm$ 0.07} & \colorbox{cyan}{1.00 $\pm$ 0.00} & \colorbox{cyan}{0.99 $\pm$ 0.01} & \colorbox{cyan}{0.99 $\pm$ 0.03}  & \colorbox{cyan}{0.98 $\pm$ 0.08} &     \\
\bottomrule
\end{tabular}%
}
\caption{Offline evaluation of BLIP-based MOSAIC engines. 
    We report Mean $\pm$ SD of the following pairwise ranking similarity measures: 
    rank-biased overlap (RBO) \raisebox{.25em}{\colorbox{pink}{ }}
    and Jaccard index (IoU) \raisebox{.25em}{\colorbox{cyan}{ }}.
}
\label{tbl:res-blip}

\end{minipage}
\end{sidewaystable}

Our results indicate that the popularity policy has a strong impact on the recommendations, 
as in all cases, the ranking similarity measures are close to $0$ with regard to other engines,
denoting no similarity and thus that all recommendations are different.
On the contrary, the representative selection policy has a rather moderate impact, 
and it is only apparent when the value of $\xi$ is very low, 
i.e., when the user tolerance to seeing diverse paintings is minimal. 
For users with high tolerance to diversity (i.e., $\xi \approx 1$) both \systemfair{CLIP}{1} and \systemfair{BLIP}{1} 
produced very similar recommendations to their respective baseline engines. 
Then, as $\xi$ decreases, the ranking similarity values tend to decrease as well, 
but only slightly and still, on average, more than 50\% of the recommendations were found to be similar;
see e.g. \systemfair{CLIP}{0.5} and \systemfair{CLIP}{0}.
A more extreme case was observed for \systemfair{BLIP}{0.5} and \systemfair{BLIP}{0}, 
where ranking similarity remained consistently high, close to 1.  

We can also observe similar results for the full-version of MOSAIC. 
Particularly, while the effect of popularity was found to be strong when the values of $\xi$ are lower, 
this engine version %(which considers the joint optimization of the three policies we have studied) 
produces similar recommendations than the baseline versions when the values of $\xi$ are higher. 
Hence, from these observations we can assume that the embeddings 
produced by the baseline multimodal engines (\system{CLIP} and \system{BLIP}) 
are already promoting paintings fairly across the story groups in the analysed paintings collection.

We performed an additional rank similarity analysis between \system{CLIP} and \system{BLIP} variants.
We observed no overlap in both IoU and RBO values, all being very close to $0$. 
This suggests that the recommendations generated with each of the multimodal backbone architectures 
are completely different, even when considering the three different policies we have studied.
From previous work, we know that and \system{BLIP} performs significantly better than \system{CLIP} on various downstream tasks, including VA RecSys, which is attributed to the modality gap effect~\cite{NEURIPS2022_702f4db7, YilmaleivaUMAP23}. Thus, as stipulated in Postulates~\autoref{post:one} and~\autoref{post:two}, we expect that BLIP-based recommendations would be perceived as being of higher quality for users. We expect the contrary to happen with the introduction of the popularity factor, which causes the recommendations to be completely different from the baseline engines,
thereby compensating for the aforementioned modality gap effect. 

Overall, from this offline evaluation we discovered that the effect of representative selection (policy 3) is minimal but slightly higher on \system{CLIP} than on \system{BLIP} engines, and that the effect of popularity (policy 2) is strong for both baseline engines. Therefore, to assess these observed effects on the quality of recommendations from a user-centric perspective, we conducted a user study described in the following section.

%% =============================================================================
\section{User study}
\label{sec:eval2}
%% =============================================================================

We conducted a subsequent user study to understand the users' perception 
towards the quality of our MOSAIC engines.
The study was approved by the Ethics Review Panel of the University of Luxembourg with ID 22-031. 

Building upon our offline evaluation, we know that representativeness has little impact on the baseline engines.
We also know that the recommendations provided by the full-version of MOSAIC 
and the popularity-based engines are substantially different,
but it is unclear how they would be perceived by end users. 
Therefore, a user study is ultimately required. 
Hence, based on the results from the offline evaluation, 
we selected the four engines that generated substantially different recommendations for our user study: 
CLIP-M and BLIP-M (reflecting the combined effect of popularity and representativeness) 
as well as the popularity-only variants CLIP-P and BLIP-P.

\subsection{Dataset}
We used the same paintings dataset as in the offline evaluation.
Please refer to \autoref{sec:dataset} for more details.

\subsection{Apparatus}
\begin{figure*}[!ht]
    \centering

    \def\w{\textwidth}
    \includegraphics[width=\w]{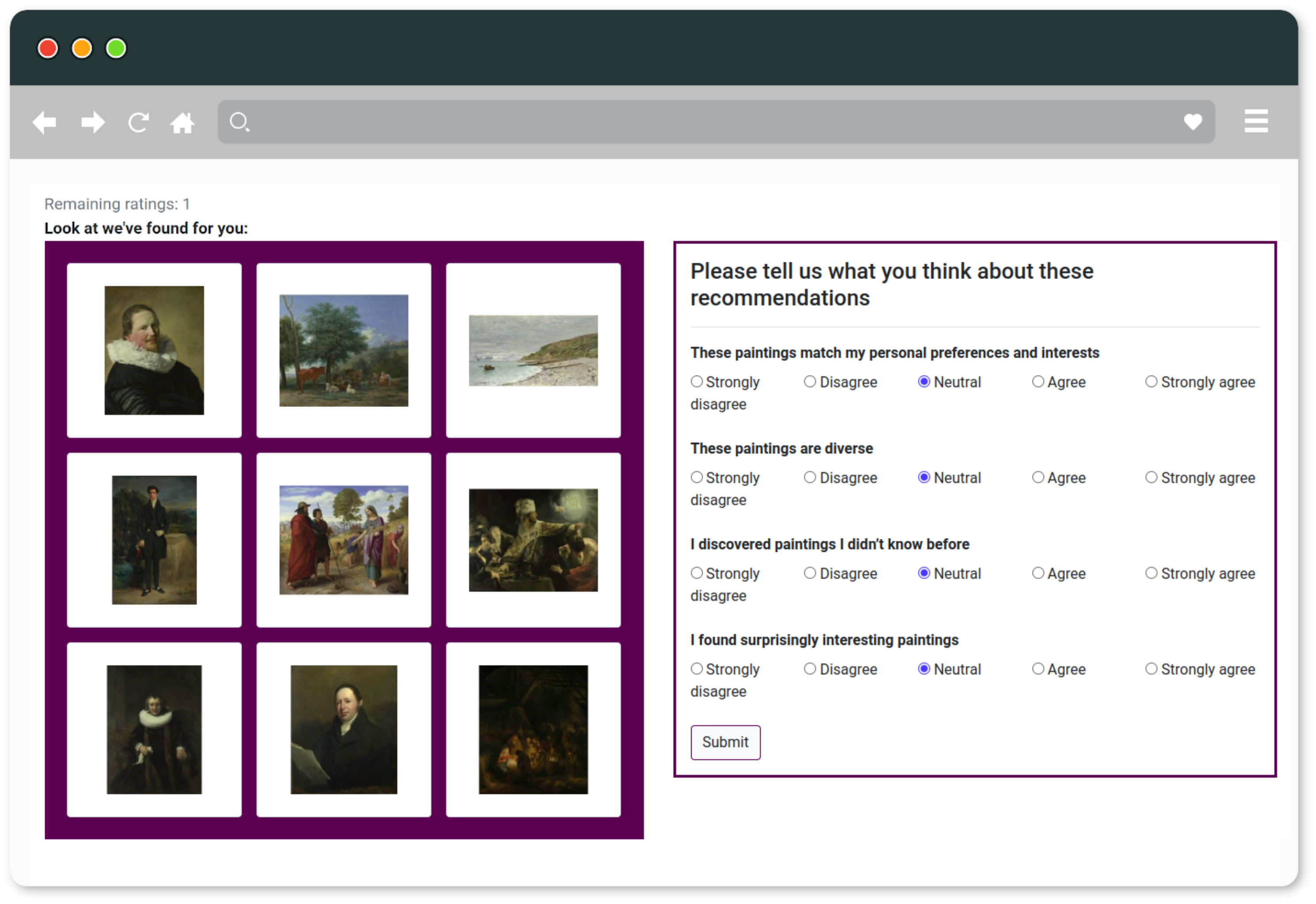}

    \caption{
        Screenshot of our web application for evaluation. %in this case shown in a desktop browser.
    }
    \label{fig:webapp}
\end{figure*}

We created a web application that collected preference elicitation ratings from users,
together with their tolerance to receiving popular and diverse paintings.
The application then retrieved a set of VA recommendations that users had to assess.
As shown in \autoref{fig:webapp}, participants were provided with one set of recommendations from each VA RecSys engine at a time.

\subsection{Participants}

We recruited a representative pool of $N=100$ crowdworkers
via the Prolific crowdsourcing platform.\footnote{\url{https://www.prolific.co/}}, according to the following eligibility criteria: 
being fluent in English, 
art is listed among their interests/hobbies, 
minimum approval rate of 100\% in previous crowdsourcing studies in the platform, 
and being active for over one year in the platform.

Our recruited participants (66 female, 34 male) were aged 27.24 years (SD=7.75) and could complete the study only once. 
Most of them had Portuguese (22\%), South African (20\%), or Mexican (14\%) nationality. 
The study took 4.9\,min on average to complete (SD=3.8) and participants were paid an equivalent hourly wage of \$15/h. 

\subsection{Procedure}

Participants accessed our web application and entered their demographic information (age, gender) on a welcome screen.
There, they were informed about the purpose of the study and the data collection policy.
Participants also indicated their tolerance to receiving popular and diverse paintings,
both in a 1--5 scale that was later normalised to the [0,1] range according to MOSAIC's formulation.
Lower values denote less user tolerance.

Then, participants advanced to the preference elicitation screen,
where they were shown one painting at random from each of the nine curated story groups.
They rated each painting in a 5-point numerical scale (5 is better, i.e. the user likes the painting the most).
Finally, users advanced to the assessment screen,
where they were shown a set of nine painting recommendations drawn from each VA RecSys engine  (the order of the engines was randomized for each user) and were asked to rate each recommended set in a 5-point Likert scale (1: Strongly disagree, ..., 5: Strongly agree).

\subsection{Design}

Participants were exposed to all engines once (within-subjects design).
Our dependent variables are widely accepted proxies of recommendation quality~\cite{pu2011user}:
\begin{description}
    \item[Accuracy:] The paintings match my personal preferences and interests.
    \item[Diversity:] The paintings are diverse.
    \item[Novelty:] I discovered paintings I did not know before.
    \item[Serendipity:] I found surprisingly interesting paintings.
\end{description}

We duplicated one of the engines (chosen at random for each user) 
to serve as an attention check and to ensure a high intra-rater agreement.
That is, if a user sees the same set of recommended paintings twice, they should rate it as much similar as before. The attention check was shown at random for each user.
Based on this rationale, we excluded 5 participants who deviated for more than 2 points in at least one of the four dependent variables.

\subsection{Results}

We investigated whether there were differences between any of the four studied MOSAIC engines,
for which we used a linear mixed-effects (LME) model where each dependent variable is explained by each engine.
User's tolerance to popular and diverse paintings are considered interaction effects (model covariates), 
and participants are considered random effects.
An LME model is appropriate here because the dependent variables are discrete and have a natural order.
In addition, LME models are quite robust to violations of several distributional assumptions~\cite{Schielzeth20}. 

We fitted the LME models (one model per dependent variable) 
and computed the estimated marginal means for specified factors.
We then ran pairwise comparisons (also known as \emph{contrasts})
with Bonferroni-Holm correction to guard against multiple comparisons.

\begin{figure}[!ht]
    \centering
    \def\w{0.4\linewidth}
    \subfloat[Accuracy]{\includegraphics[width=\w]{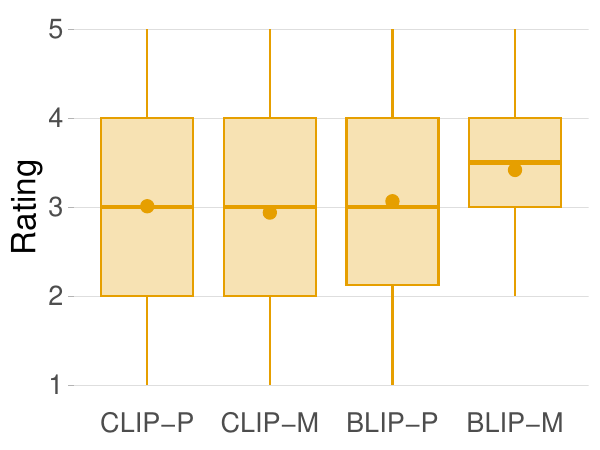}} \hfill
    \subfloat[Diversity]{\includegraphics[width=\w]{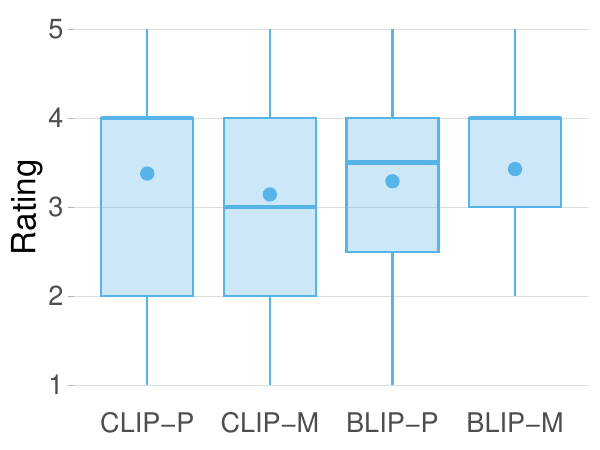}} \hfill
    \subfloat[Novelty]{\includegraphics[width=\w]{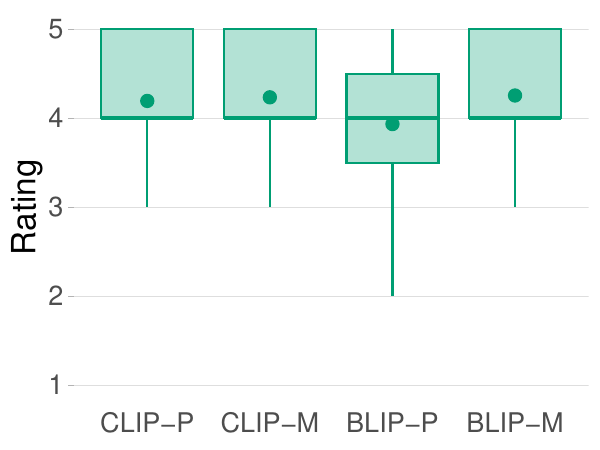}} \hfill
    \subfloat[Serendipity]{\includegraphics[width=\w]{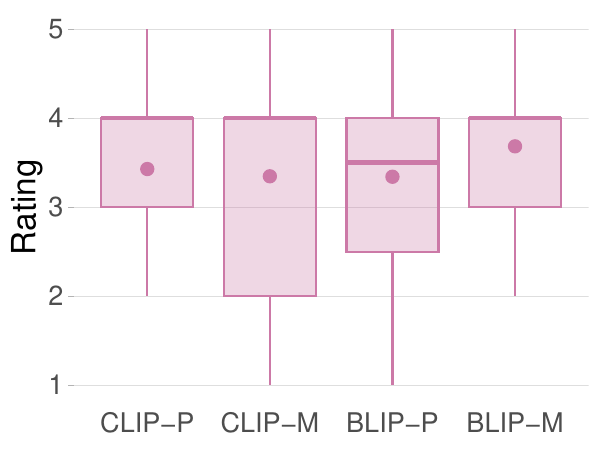}}

    \caption{Distribution (box plots) of user ratings. Dots denote mean values.}
    \label{fig:prolific-ratings-overall}
\end{figure}

\autoref{fig:prolific-ratings-overall} shows the distributions of user ratings. 
We report below statistical significance ($p$-values) and effect sises ($r$)
to better gauge the differences between our four MOSAIC engines.

\paragraph{Accuracy analysis}
Differences were statistically significant between BLIP-P and BLIP-M ($p < .01, r=0.18$)
and between CLIP-P and BLIP-M ($p < .01, r=0.18$).
Differences between CLIP-M and BLIP-M were also statistically significant ($p < .01, r=0.21$).
All other comparisons were not found to be statistically significant.
Effect sizes suggest a small to moderate practical importance of the results.

\paragraph{Diversity analysis}
No statistically significant differences were found between engines ($p > .05$).
The largest effect size was $r=0.13$ when comparing CLIP-M and BLIP-M,
suggesting therefore a small practical importance of the results.

\paragraph{Novelty analysis}
Differences between BLIP-P and all the other engines were found to be statistically significant, i.e., 
BLIP-P vs. CLIP-P ($p < .01, r=0.17$),
BLIP-P vs. BLIP-M ($p < .001, r=0.21$),
and BLIP-P vs. CLIP-M ($p < .001, r=0.22$).
All other comparisons were not found to be statistically significant.
Effect sizes suggest a small to moderate practical importance of the results.

\paragraph{Serendipity analysis}
Differences were statistically significant between BLIP-P and BLIP-M ($p < .05, r=0.15$)
and between CLIP-M and BLIP-M ($p < .05, r=0.16$).
All other comparisons were not found to be statistically significant.
Effect sizes suggest a small practical importance.

\paragraph{Ranking similarity analysis}
We conducted an additional analysis that checked whether our participants did receive truly personalised recommendations.
For this, we computed RBO and IoU in a pairwise manner among all users exposed to the same engine.
Tables~\ref{tab:ranking_overlap} and \ref{tab:joint_ranking_overlap} present the results of this analysis.
As shown in the tables, there is little overlap in the rankings produced by each engine.
This analysis thus indicates that each participant indeed was shown a personalisd set of recommendations
and that each engine produced very different rankings.

\begin{table}[!h]
\centering
\small
\caption{Within-ranking overlap results, showing Mean $\pm$ SD of IoU and RBO measures.}
\label{tab:ranking_overlap}
\begin{tabular}{l*4c}
\toprule &
  \textbf{CLIP-P} &
  \textbf{BLIP-P} &
  \textbf{CLIP-M} &
  \textbf{BLIP-M} \\
\midrule
  IoU &
  0.12 $\pm$ 0.76 &
  0.01 $\pm$ 0.07 &
  0.25 $\pm$ 0.26 &
  0.25 $\pm$ 0.18
\\
  RBO &
  0.01 $\pm$ 0.07 &
  0.01 $\pm$ 0.07 &
  0.18 $\pm$ 0.17 &
  0.28 $\pm$ 0.18
\\
\bottomrule
\end{tabular}
\end{table}

\begin{table}[!h]
\centering
\small
\caption{Between-ranking overlap results, showing Mean $\pm$ SD of 
    IoU \raisebox{.25em}{\colorbox{cyan}{ }} and
    RBO \raisebox{.25em}{\colorbox{pink}{ }} 
    measures.
}
\label{tab:joint_ranking_overlap}
\begin{tabular}{l*4c}
\toprule &
  \textbf{CLIP-P} &
  \textbf{BLIP-P} &
  \textbf{CLIP-M} &
  \textbf{BLIP-M}
\\
\midrule
  \textbf{CLIP-P} &
  &
  \colorbox{pink}{0.01 $\pm$ 0.16} &
  \colorbox{pink}{0.01 $\pm$ 0.04} &
  \colorbox{pink}{0.00 $\pm$ 0.01}
\\
  \textbf{BLIP-P} &
  \colorbox{cyan}{0.01 $\pm$ 0.01} &
  &
  \colorbox{pink}{0.00 $\pm$ 0.01} &
  \colorbox{pink}{0.00 $\pm$ 0.00}
\\
  \textbf{CLIP-M}  &
  \colorbox{cyan}{0.01 $\pm$ 0.01} &
  \colorbox{cyan}{0.00 $\pm$ 0.01} &
  &
  \colorbox{pink}{0.00 $\pm$ 0.01}
\\
  \textbf{BLIP-M}  &
  \colorbox{cyan}{0.00 $\pm$ 0.01} &
  \colorbox{cyan}{0.00 $\pm$ 0.01} &
  \colorbox{cyan}{0.00 $\pm$ 0.01} &
\\
\bottomrule
\end{tabular}%
\end{table}

In our user study, participants indicated a high tolerance to receiving diverse recommendations, 
with a mean $\xi=0.74$ (SD=0.26). Only 3 participants indicated $\xi = 0$.
Consequently, the implementation of a representative selection policy yielded minimal impact overall, 
as seen in \autoref{fig:prolific-ratings-overall}. 
This is consistent with our offline evaluation (\autoref{subsec: offline_results}), 
where the impact of representative selection was found to be moderate 
and only apparent when the value of $\xi$ is close to zero, 
i.e., denoting minimal user tolerance to seeing diverse paintings. This was also reflected on the actual diversity of the recommendations, 
measured as the median number of different story groups covered per set of 9 paintings 
(BLIP-M = 7, CLIP-M = 6, CLIP-P = 5, BLIP-P = 6).
While BLIP-M produced the most diverse recommendations overall,
the differences between engines are not statistically significant ($p > .05$). Overall, from this user study, we can conclude that users appreciated 
the inclusion of the popularity policy in MOSAIC, 
indicating that other stakeholders such as collectors or art enthusiasts
play an important role in the curation of VA content.
According to the four user-centric measures considered, 
BLIP-M was consistently perceived as the top-performer VA RecSys engine, followed by BLIP-P.
We can conclude therefore that BLIP is the most adequate backbone engine for MOSAIC.
In the following section we discuss the implications of our findings.

%% =============================================================================
\section{Discussion}
\label{sec:disc}
%% =============================================================================
 
We reflect on the implications of our findings and derive valuable insights
on how to design next-generation VA RecSys engines blending learning, discovery, and user satisfaction.

\subsection{A Palette of Possibilities: From Personalisation to Learning and Discovery}

VA RecSys have traditionally focused on providing users with personalised recommendations based on their past preferences and behaviour. 
However, as these systems continue to evolve, there is a growing realisation that they have the potential to offer much more. 
Considering the vast and dynamic status quo in the art domain, by jointly optimising for other objectives, such as learning and discovery, 
VA RecSys can create a palette of possibilities that goes beyond simply catering to users' existing tastes. 
This transition from pure personalisation to learning and discovery is important for several reasons. 
First, it enables users to expand their knowledge and appreciation of VA beyond their existing preferences. 
Second, it also enables VA RecSys to serve as a tool for lifelong learning and personal growth, rather than simply a means of entertainment.
Further, by discovering new art styles and genres, users can gain a deeper understanding of the cultural and historical contexts in which different art forms originated, and broaden their aesthetic horizons. This can be especially important for users who may not have had much exposure to the arts in the past, and who may not be aware of the rich and diverse range of artistic traditions that exist around the world. 

Furthermore, the transition from pure personalisation to learning and discovery also has significant implications for the development of VA RecSys. 
By optimising these systems to serve a broader range of objectives, developers can create a more flexible and adaptable framework that can be applied to a wide range of use cases. For example, VA RecSys could be used in educational contexts to help students learn about different art styles and movements, or in museum settings to provide visitors with a more personalised and engaging tour experience. By expanding these possibilities beyond pure personalisation, we can unlock a whole new range of applications and use cases. Our results indicate that departing from solely optimising for user preferences to factor in additional objectives such as paintings' popularity have managed to strike a balance across all beyond-accuracy measures. Particularly, BLIP-P was mostly preferred while the other multistakeholder-aware engines also performed similarly, 
thus opening a new avenue for designing innovative policies to be jointly optimised.

\subsection{Into the Canvas: Multimodality and Semantics} 

Recent research has focused on advanced VA representation learning techniques to capture the semantic relatedness of artworks~\cite{Yilmaleiva23, messina2020curatornet, he2016vista}. 
This is crucial for VA RecSys to derive recommendations that preserve some notion of art semantics. 
In this regard, multimodal data, which includes both visual and textual information, has become increasingly important.
By analysing both image content and accompanying text, VA RecSys can better understand the context and meaning of an artwork, which in turn allows for more relevant recommendations. 
Multimodal representation learning techniques 
have shown great potential in terms of capturing latent semantics that are not obvious to the naked eye.  

Particularly the state-of-the-art architectures CLIP and BLIP have undoubtedly demonstrated notable success in VA RecSys. However, a pressing question looms: to what extent and depth do they truly capture the rich semantics embedded within visual art?

Especially since the semantics of artwork is a broader notion that encapsulates various perspectives from an array of stakeholders in the art ecosystem and concepts beyond mere aesthetic elements depicted in paintings or written in their textual descriptions.  Our user study shows that introducing the elements of popularity or representativeness to either CLIP or BLIP achieved results on par with their respective baselines. This signifies that, while multimodal approaches present a promising foundation, it becomes evident that comprehending the semantic depth of artwork transcends the modalities and therefore requires looking ``beyond the canvas''. This opens up exciting research avenues for designing more comprehensive VA RecSys policies that not only accommodate the diverse perspectives of relevant stakeholders but also pave the way for a deeper understanding of the intricate world of visual arts. 

Furthermore, in our user study, we observed that BLIP-P was preferred over CLIP-P 
and BLIP-M was preferred over CLIP-M, 
confirming that recommendations powered by BLIP are perceived to be better than CLIP. 
This indicates that the two backbone architectures allow different levels of exposition to VA semantics. 
This adheres to the findings about the modality gap effect reported in previous works, 
which validates Postulate~\autoref{post:one}. 
Additionally, CLIP-P outperformed CLIP-M which produced similar recommendations with the baseline CLIP as informed by our offline evaluation, suggesting that the popularity policy indeed compensated for the modality gap effect. However, it was not enough to outperform BLIP therefore partially validating Postulate~\autoref{post:two}. Nonetheless, our results indicate a potential for addressing the modality gap effect in CLIP by introducing joint optimisation objectives.

\subsection{Beyond the Canvas: Navigating the Art Landscape}
% Multistakeholders, Why does popularity matter and who is responsible?

While VA RecSys is primarily intended to personalise recommendations to individual users, 
there has been a missed opportunity to navigate the larger art ecosystem which includes a number of stakeholders. 
For example, art collectors, who are often interested in acquiring unique or rare pieces of art, 
can drive up the demand and value of certain works, thereby making certain artists or artworks more popular. 
Popularity is therefore an important aspect in VA RecSys,  which has not been sufficiently explored in the research literature. 
The results from our user study indicate that factoring in popularity was perceived by users to improve the quality of recommendations, 
not only in terms of accuracy but also in terms of serendipity and diversity. 
This highlights the need for further investigation in this regard, 
especially since there are many factors and stakeholders that contribute to the popularity of a painting. 
Furthermore, their historical significance, artistic merit, and cultural relevance also play a vital role. 
In this work, we conceptualise popularity as an influential factor shaping user preferences, driven by crowd behaviour, which represents a significant stakeholder contributing substantially to the context of VA RecSys. Future work can also explore integrating various crowd objectives and facets into the MOSAIC framework, enhancing the framework's ability to efficiently represent stakeholders.  Hence, we are confident that our findings will highlight the relevance of multistakeholder-awareness in VA RecSys. 

\subsection{Looking Ahead: Limitations and Future work}

One limitation of our experiments is that, despite the large number of participants,
we have considered a relatively small set of elicited preferences (ratings of only 9 paintings per user). 
Preference elicitation is a longstanding challenge in RecSys, 
as collecting more user preferences usually improves recommendation quality, 
but this is not always feasible or accessible. 
Although our results confirm the relevance of MOSAIC in tackling the elicitation challenge,  
more studies on VA RecSys performance with varying preference data are needed.
Another limitation is that we only considered two stakeholders to optimise for, 
in addition to the user preferences,
and there may be others to cater for learning and discovery.
This is nonetheless an exciting opportunity for future work.

Our experiments show that using multimodal data leads to significant improvements in recommendation quality
and that the effect of the modality gap in CLIP can potentially be minimised by introducing joint optimisation policies.
However, further research is needed to determine the extent to which these findings will be observed in other domains.
Future work should also consider more stakeholders, in order to provide a more nuanced understanding of the art ecosystem.

%% =============================================================================
\section{Conclusion}
\label{sec:conclusion}
%% =============================================================================

We have explored the challenging task of visual art recommendation, 
taking into account subjective user experiences as well as the input of relevant stakeholders 
such as museum curators and crowds of visitors. 
We have proposed MOSAIC, a novel multimodal multistakeholder-aware approach 
that goes beyond traditional accuracy measures and promotes user-centric measures 
such as novelty, serendipity, and diversity. 
Our results showed that departing from solely optimising for user preferences 
to factor in multistakeholder awareness has a positive impact in promoting learning and discovery. 
These findings suggest that MOSAIC has the potential to add value not only for users, 
but also for other stakeholders, including artists, curators, and art organizations. 
As the field of VA RecSys continues to evolve, MOSAIC represents the first cornerstone 
to navigate the complex landscape of the art ecosystem. 

\bmhead{Acknowledgements}
This work is supported by the Horizon 2020 FET program of the European Union through the ERA-NET Cofund funding (project BANANA, grant CHIST-ERA-20-BCI-001) and Horizon Europe's European Innovation Council through the Pathfinder program (SYMBIOTIK project, grant 101071147).

\bibliography{sn-bibliography}

\end{document}